\documentclass[showpacs,aps,prf,onecolumn,superscriptaddress,preprint]{revtex4}

\usepackage{graphicx}
\usepackage{amsmath}
\usepackage{amssymb}
\usepackage[sort&compress]{natbib}
\usepackage{color}
\usepackage{CJK}
\usepackage{dcolumn}   
\usepackage{harpoon}

\begin{document}

\title{Influence of grain bidispersity on dense granular flow \\ in a two dimensional hopper}

\author{Changhao Li} \author{Xin Li} \author{Tengfei Jiao} \author{Fenglan Hu} \author{Min Sun}
\author{Decai Huang}
\email[Corresponding author: ] {hdc@njust.edu.cn}
\affiliation{Department of Applied Physics, Nanjing University of Science and Technology, Nanjing 210094, China}

\date{\today}

\begin{abstract}

  Discrete element method is conducted to investigate the bidisperse dense granular flow having big and small grains in a two dimensional hopper. A half-circular dynamical force arch is observed above the outlet and Beverloo's law is verified to describe the relationship between flow rate and outlet size. The bidisperse flow can reach the maximum flow rate when a small fraction of big grains is added into the monodisperse flow with only small grains. The distributions of contact force, packing fraction and grain velocity indicate that the flow properties in the hopper are closely related to the local flow characteristics of the key area which almost overlaps with the force arch. The interior packing structures, i.e., ordered arrangement for monodisperse flow and disordered arrangement for bidisperse flow, play a dominant role in the flow pattern transition from mass flow to funnel flow. The earlier occurrence of the transition can effectively increase the grain velocity and therefore improve the flow rate.

\end{abstract}


\maketitle
\section{Introduction}

Dense granular flow is often encountered in natural environments and industrial productions and has attracted extensive interest in the scientific and engineering fields\cite{Jaeger1996,Aranson2006,Iverson1997}. Complex phenomena, such as mass-to-funnel flow pattern transition\cite{PT22P243,RobertsCES2002,WojcikPT2012,ZhangPT2018,JiShunyingPT2019,JiShunyingSM2020}, mixing and segregation\cite{HuangEPJE2013,ShiPRE2007,WangCPB2017}, and avalanches\cite{Bursik2005,Denisov2016} have been widely investigated. The dense flow with bidisperse grains in a hopper is an important research topic because of its ubiquity. Understanding of the whole flow rate and interior flow behavior is of primary importance in ensuring reproducible and efficient handling in industrial operations.

The quantitative model for the dense flow rate of monodisperse grains in a hopper was first proposed by Hagen\cite{Hagen1852} and rediscovered by Beverloo et al.\cite{Beverloo1961}. In this model, the outlet is located at the center of the hopper. The relationship between flow rate and outlet size takes the following form:
\begin{equation}
Q=C{\sqrt{g^{*}}}(D-kd)^{n-1/2},
\label{BeverlooEq1}
\end{equation}

\noindent where $n$ is the spatial dimension, i.e, $n$ = 2 and 3 for 2D and 3D respectively. $D$ is the outlet size. $d$ is the grain diameter. $g^*$ is the effective gravity acceleration. $C$ is a fitting dimensional coefficient that is possibly related to factors such as friction coefficient and bulk density. $k$ is regarded being resulted from the empty annulus effect related to the geometric shape of the grain. The free-fall arch hypothesis assumes that the grains are closely compacted above the outlet and the contact force has a half-circular structure. The grains fall freely with zero initial velocity from the force arch region. Beverloo's formula has been verified for the past few decades\cite{NeddermanCES1982,Zuriguel2012PRL108P248001,AguirrePRL2010,ZhengPT345P676Y2019,BenyaminePRE2014,HuangPT2021,ArteagaCES1990,HumbyCES1998}.

When the flow comprises bidisperse mixtures with big and small grains, Arteaga and T\"{u}z\"{u}n modified Beverloo's equation as follows\cite{ArteagaCES1990,HumbyCES1998}:
\begin{equation}
Q=C_{\rm m}{\sqrt{g^{*}}}(D-k_{\rm m}d_{\rm m})^{n-1/2},
\label{BeverlooEq2}
\end{equation}

\noindent where the subscript $\rm m$ represents the bidisperse mixtures. $d_{\rm m}=X_{\rm B}d_{\rm B}+(1-X_{\rm B})d_{\rm S}$ is the mean diameter. $d_{\rm B}$ and $d_{\rm S}$ are the diameters of big and small grains. The mixing index $X_{\rm B}$ is defined by the mass ratio $X_{\rm B}={\sum m_{\rm B}}/{\sum (m_{\rm B}+m_{\rm S})}$. $m_{\rm B}$ and $m_{\rm S}$ are the masses of single big and small grains. $X_{\rm B}=0$ and $1.0$ represent the monodisperse flow containing only small and big grains. The others represent the bidisperse flow that mixes big and small grains. The experimental results show that the difference between grain diameters greatly affects the flow rate and local flow behavior. Coarse and fine continuous flows are defined when the flow is dominated by big and small grains respectively. Dilatancy occurs as a result of the grain's flowing, which in turn results in a lower dynamical packing fraction compared with that observed in the initial static packing case. The flow rate decreases along with an increasing mixing index, and the maximum dynamical packing fraction appears at a moderate mixing index. The dispersity of grains results in empty annuluses $k_{\rm m}=1.85$ and $1.40$ for the coarse and fine continuous flows, respectively.

Benyamine and Zhou et al\cite{BenyaminePRE2014,ZhouPRE2015} replicated the bidisperse flow in a hopper via experiments and simulations, in which the outlet was formed by two plat baffles. Their findings reveal that the bidisperse flow has a higher dynamical packing fraction compared with the monodisperse case. However, they reported the maximum flow rate at a moderate mixing index. Such discrepancy might be ascribed to the grain shape, which takes the form of smooth glass spheres in their experiments. Although they found no fall arch, the flow behavior was still gradually affected by the local flow properties in the region, especially around the outlet. These results indicate that both grain velocity and packing fraction have self-similarity characteristics. Meanwhile, the expressions proposed by Beverloo\cite{Beverloo1961} and Janda\cite{Zuriguel2012PRL108P248001} equally fit the relationships between flow rate and outlet size. To obtain further insights into the influence of grain dispersity on the flow kinetic properties, the bidisperse flow with big and small grains warrants intensive investigation.

In this paper, numerical simulations are performed to explore the bidisperse flow in a two dimensional(2D) hopper. The simulation method is first described in Sec. \ref{SimMod}. The key area and corresponding calculation of the flow rate are also introduced. In Sec. \ref{Results}, the dependence of the flow rate on outlet size is tested by using Beverloo's formula. The influence of size and mass ratios on flow rate is repeatedly examined. Afterwards, the effect of grain dispersity on the distribution of packing fraction, grain velocity, and contact force is examined. The transition from mass flow to funnel flow patterns are then discussed for both monodisperse and bidisperse flows. The major results are summarized in the conclusion.

\section{Simulation method}
\label{SimMod}

\begin{figure}[htbp] 
\centering
\includegraphics[width=7 cm,trim=95 220 100 200,clip]{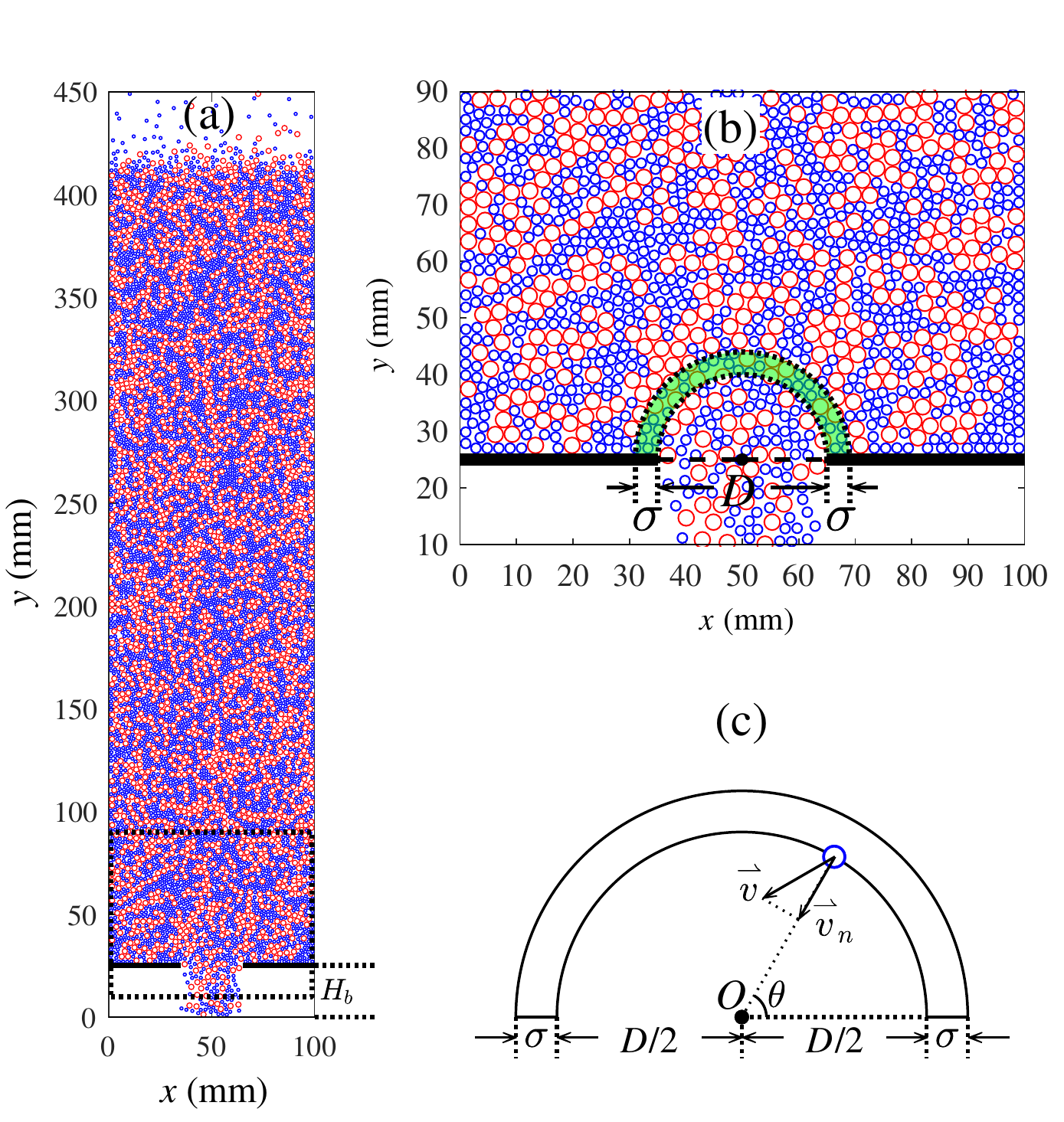}
  \caption{(Color online). (a) Snapshot of the simulation system. (b) The enlarged part encircled by dotted lines in (a), where the key area is marked by dotted lines filled in green. The width of the key area is set as $\sigma=2d_{\rm S}$. The inner diameter is equivalent to the outlet size $D$. (c) Calculation of packing fraction and flow rate. $\mathop{v}\limits^{\rightharpoonup}$ and ${\mathop{v}\limits^{\rightharpoonup}}_n$ are the grain velocity and the corresponding normal velocity respectively.}
\label{fig:FigModelKey}
\end{figure}

Discrete element method(DEM) is applied to investigate the granular flow in a quasi-2D hopper as shown in Fig.\ref{fig:FigModelKey}(a). The rectangular hopper comprises two walls with height $H=450~{\rm mm}$, width $W=100~{\rm mm}$, and thickness $T=2~{\rm mm}$. A central outlet formed by two plat baffles is located at height $H_{\rm b}=25.0~{\rm mm}$. The size of the outlet $D$ can be changed in the simulations. The grains are composed of circular disks and have the same thickness as that of the hopper.

In the simulations, the particle motion is described using Newton's equations similar to our previous works\cite{Huang2006PRE,Huang2011PLA,Huang2018CPB,HuangPT2021}. The effective gravity acceleration is set as $g^{*}=g{\sin}18^{\circ}~{\rm m/s^{2}}$. The Verlet algorithm is used to update the positions and velocities of the grains in each simulation time step. The translational motion in the hopper plane and the rotational motion along to the axis perpendicular to the hopper plane are both considered. The interaction between two contacted grains is treated as soft spheres, in which the forces along the normal and tangential directions are considered. The normal interaction is a function of the overlap at the contacting point modeled by using the Cundall-Strack formula \cite{Cundall1979,Schafer1996}.
\begin{equation}
  F_{n}={\frac{4}{3}}E^{*}{\sqrt{R^{*}}}{\delta_{n}}^{3/2}
       -2.0{\sqrt{\frac{5}{6}}}{\beta}{\sqrt{S_{n}m^{*}}}V_{n},
  \label{CundallFn}
\end{equation}

The tangential component is taken as the minor tangential force with a memory effect and the dynamic frictional force.
\begin{equation}
  F_{\tau}=-\min({S_{\tau} {\delta}_{\tau}}-2.0{\sqrt{\frac{5}{6}}}{\beta}{\sqrt{S_{\tau}m^{*}}}V_{\tau}, {{\mu} F_{n}}){\rm{sign}}({\delta}_{\tau}),
  \label{CundallFt}
\end{equation}

In Eqs.(\ref{CundallFn})(\ref{CundallFt}), $n$ and $\tau$ represent the normal and tangential directions at the contact point, respectively. $\delta_{n}$ and $\delta_{\tau}$ denote the displacements that take place since the time $t_0$ when the contact is first established. The detailed values can be calculated as follows:

\vspace{10 pt} 
\noindent
$\beta={\frac{{\rm ln}e}{{\sqrt{{\rm ln}^{2}e}+{\pi}^{2}}}}$, ${\frac{1}{m^{*}}}={\frac{1}{m_{i}}}+{\frac{1}{m_{j}}}$ \\

\noindent
$S_{n}=2E^{*}{\sqrt{R^{*}{\delta_{n}}}}$,  $S_{\tau}=8E^{*}{\sqrt{R^{*}{\delta_{n}}}}$,\\

\noindent
$E^{*}={\frac{1-{\nu_{i}^{2}}}{E_{i}}}+{\frac{1-{\nu_{j}^{2}}}{E_{j}}}$,
${\frac{1}{R^{*}}}={\frac{1}{R_{i}}}+{\frac{1}{R_{j}}}$ \\

\noindent where $e$ is the restitution coefficient. $m_i$ and $m_j$ are the masses of the contacting grain $i$ and grain $j$. $S_{n}$, $S_{\tau}$ characterize the normal and tangential stiffness of the grains. $E$ and ${\nu}$ are the Young's modulus and Poisson's ratio, respectively. The friction coefficient $\mu $ is fixed in our simulations. The collision between the grain and wall is treated as the grain-grain collision, except that the wall has infinite mass and diameter. The detailed values are presented in Table 1.
\begin{center}
{\bf{Table 1}} The parameters of the grains \\
\begin{tabular} { p{5.8cm} p{1.45cm} p{0.95cm} }
   \hline
   Quantity & Symbol   & Value \\
   \hline
   Diameter of small grain (mm) & \textit {$~~~d_{\rm S}$} & 2.0 \\
   Density ($\rm {10^{3}kg/m}^{3}$) & \textit {$~~~\rho$} & 7.8 \\
   Young modulus (GPa) & \textit {~~~E} & 1.0 \\
   Poison ratio & $~~~\nu $ & 0.3 \\
   Friction coefficient & $~~~\mu$ & 0.5 \\
   Simulation time-step (s) & $~~~dt$ & $10^{-6}$ \\
   \hline
\end{tabular}
\label{TableGrainPara}
\end{center}

At the beginning of each simulation, the outlet is closed and the hopper is completely empty. The grains are randomly poured into the hopper and fall down under the self-gravity. The initial piling height is about $H_p \approx 425~{\rm mm}$. Afterward, the outlet is opened and the flow is triggered. The periodical condition is adopted to maitain a constant packing height, which means that the gains re-enter the hopper from the top after passing through its bottom. In the simulations, the small grain diameter is fixed at $d_{\rm S}=2.0~{\rm mm}$. The big to small grain diameter ratio is defined as $\gamma=d_{\rm B}/d_{\rm S}$.

Our previous works show that the flow properties in a key area are crucial to both the flow state and granular outflow rate\cite{Huang2006PRE,Huang2011PLA,HuangPT2021,Huang2007CTP}. The same key area is selected as shown in Fig.\ref{fig:FigModelKey} (b) denoted by the dotted lines. The width of the key area is set to $\sigma=2d_{\rm S}$. The local packing fraction can be directly calculated from the ratio of volume occupied by grains in the key area to the whole volume of the key area. Therefore, the local packing fraction and flow rate are both directly calculated.
\begin{equation}
 \begin{aligned}
  & \rho={\frac{ {\sum_i {A_i}} } { S_{\rm C} } } \\
  & Q={\frac{{\sum_i {q^{n}_i}}} {\sigma} }.
 \end{aligned}
 \label{PackFlowQ}
\end{equation}
\noindent where $\sum_i$ represents the summation of all grains in the key area.  As shown in Fig.\ref{fig:FigModelKey}(c), ${q^{n}_i}={\frac{A_{i}}{A_{g}}v^{n}_{i}}$ is the flow rate of grain $i$ contributing to the key area. $A_g$ is the grain area, and $A_{i}$ is the intersection between the grain and the key area. $v^{n}_{i}$ is the normal velocity relative to the inner circle with the size of the left or right outlet as shown in Fig.\ref{fig:FigModelKey}(c). $S_{\rm C}$ is the area of the key area. According to the conservation law of mass, the local flow rate of the key area is equal to that of the outlet in a steady flow state.

\section{RESULTS AND DISCUSSIONS}
\label{Results}

In Beverloo's arguments, the dense flow rate is determined based on the hypothesized free-fall arch above the outlet. For the analysis, spatial maps of the time-averaged contact force between grains are plotted for three mixing indexes, namely $X_{\rm B}=0,~0.5$ and $1.0$. The diameters of the small and big grains are set as $d_{\rm S}=2.0$ and $d_{\rm B}=2.4~\rm{mm}$. A half-circular dynamical force arch is clearly observed above the outlet labeled by the contour solid lines $35.0$ and $50.0$. The key area is marked by green dotted lines. The shape of the force arch almost overlaps with that of the key area. A strong contact force is observed for those grains above the force arch, which indicates that these grains are closely crowded. When these grains continue flowing toward the outlet, the contact force becomes weaker, and the half-circular arch gradually turns into a flat line labeled by $10.0$. In other words, the squeezed grains become loose and fall out of the hopper under the effects of self-gravity.
\begin{figure}[htbp]
\centering
\includegraphics[width=8 cm,trim=120 300 125 300,clip]{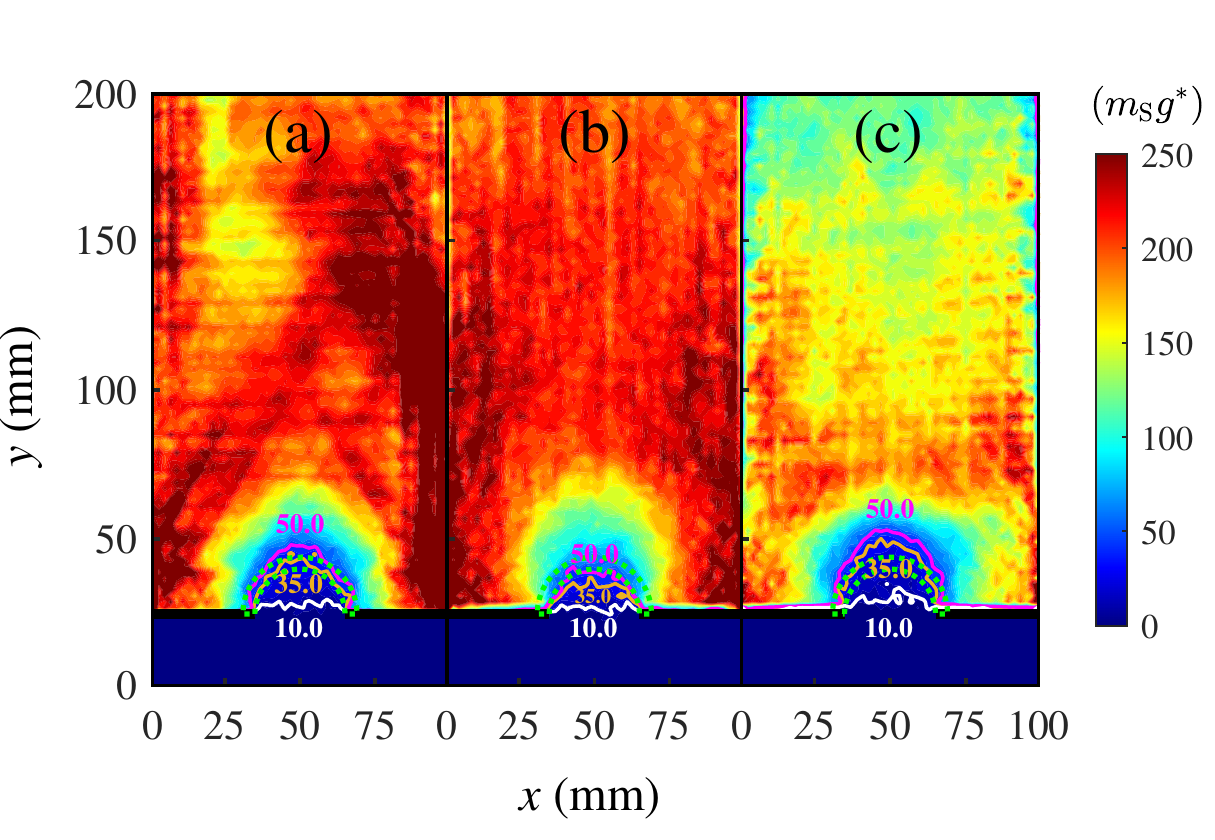}
  \caption{(Color online). Distribution of the contact force between grains for three mixing indexes: (a) $X_{\rm B}=0.0$, (b) $X_{\rm B}=0.5$ and (c) $X_{\rm B}=1.0$. The diameters of the small and big grains are $2.0$ and $2.4~{\rm mm}$. The outlet size is set as $D=30.0~{\rm mm}$. The contact force is normalized with respect to the gravity of small grain (${m_{\rm S}}g^*$). The brown ($35.0$) and purple ($50.0$) solid lines indicate the half-circular dynamical force arch above the outlet. The key area is denoted by green dashed lines.}
\label{fig:FigForceDist}
\end{figure}

\begin{figure}[htbp]
\centering
\includegraphics[width=7 cm,trim=135 290 140 290,clip]{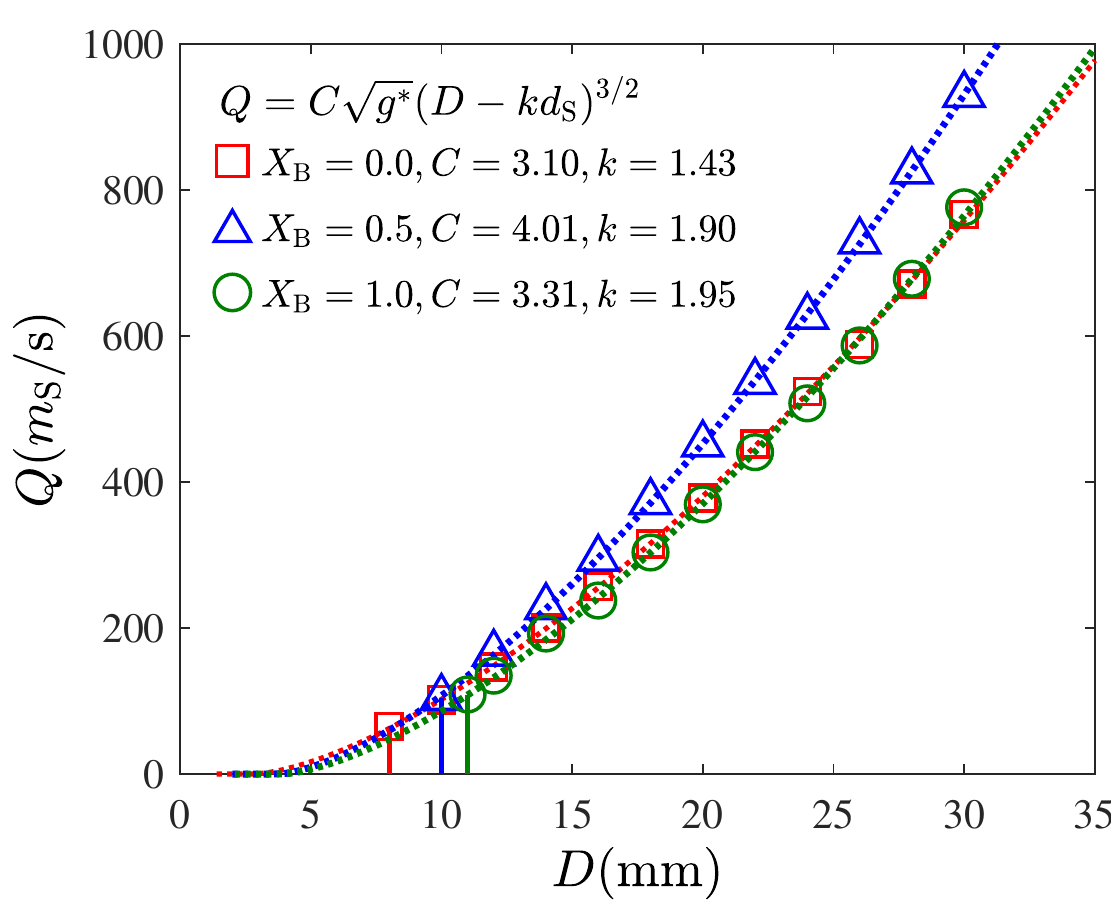}
  \caption{(Color online). Simulated results of flow rate as a function of outlet size. The flow rate is normalized by small grain ($m_{\rm S}$). The squares, triangles and circles denote the mass ratio $X_{\rm B}=0,0.5$ and $1.0$. The diameter ratio is fixed at $\gamma=1.2$. The dotted lines are obtained by using Beverloo's equation. The solid lines indicate the locations of the clogging in our simulations.}
\label{fig:FigBeverloo}
\end{figure}

An interesting point from Fig.\ref{fig:FigForceDist} is that the force arch labeled by $35.0$ and $50.0$ for $X_{\rm B}=0.5$ is obviously smaller than that for $X_{\rm B}=0$ and $1.0$. Following the free-fall arch hypothesis, a larger flow rate corresponds to a larger force arch. To test this prediction, Fig.\ref{fig:FigBeverloo} plots the flow rate $Q$ of the key area as a function of outlet size $D$ for the same flow conditions used in Fig.\ref{fig:FigForceDist}. Simulation results coincide well with Beverloo's equation in the power law of 3/2. The fitting parameter $k$ is equivalent to $1.43$ and $1.95$ for the monodisperse flow with small $(X_{\rm {B}}=0.0)$ and big $(X_{\rm {B}}=1.0)$ grains, respectively. The flow with bidisperse grains $(X_{\rm {B}}=0.5)$ has moderate fitting parameters $k=1.90$, which is consistent with the free-fall arch hypothesis. A bigger grain has a larger empty annulus, which results in a smaller effective outlet size and flow rate. However, Fig.\ref{fig:FigBeverloo} shows that the flow rate at the bidisperse case of $X_{\rm B}=0.5$ is larger than that at the monodisperse cases of $X_{\rm B}=0$ and $1.0$, which contradicts to free-fall arch hypothesis. In addition, colgging is frequently observed at the outlet size of $D/d \leq 4.5$ for the monodisperse flow in our simulations, and the clogging for the bidisperse flow tends to happen at a moderate outlet size. This finding agrees with the observations of Benyamine\cite{BenyaminePRE2014}.
\begin{figure}[htbp]
  \centering
  \includegraphics[width=7 cm,trim=135 290 140 265,clip]{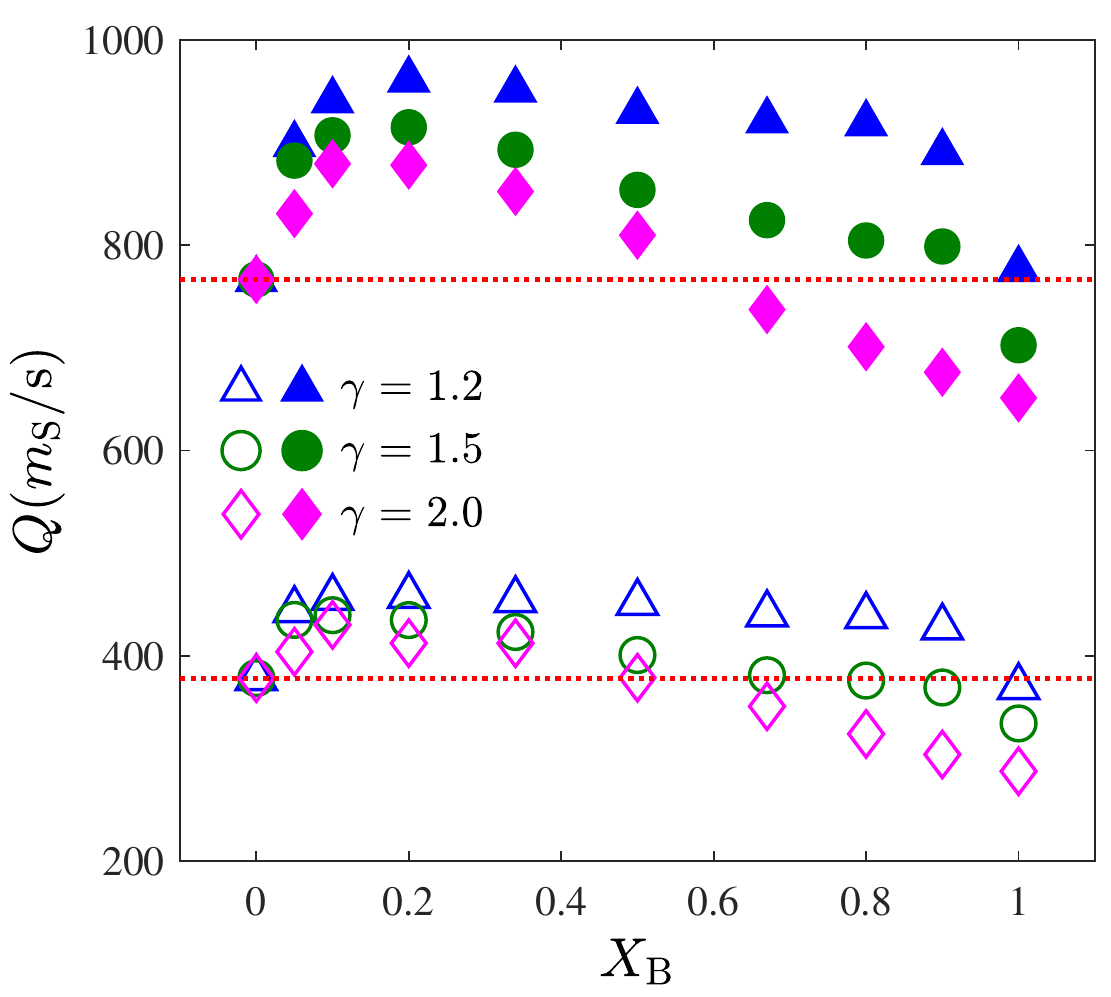}
  \caption{(Color online). Simulated results for the relationship between flow rate and mixing index. The flow rate is expressed in small grain mass ($m_{\rm S}$) per second. The open and solid symbols correspond to outlet sizes of $D=20~{\rm mm}$ and $30~{\rm mm}$, respectively. The dotted lines denote the monodisperse flow rate $X_{\rm B}=0.0$.}
\label{fig:FigQoutXB}
\end{figure}

\begin{figure}[htbp]
  \centering
  \includegraphics[width=7 cm,trim=120 120 120 120,clip]{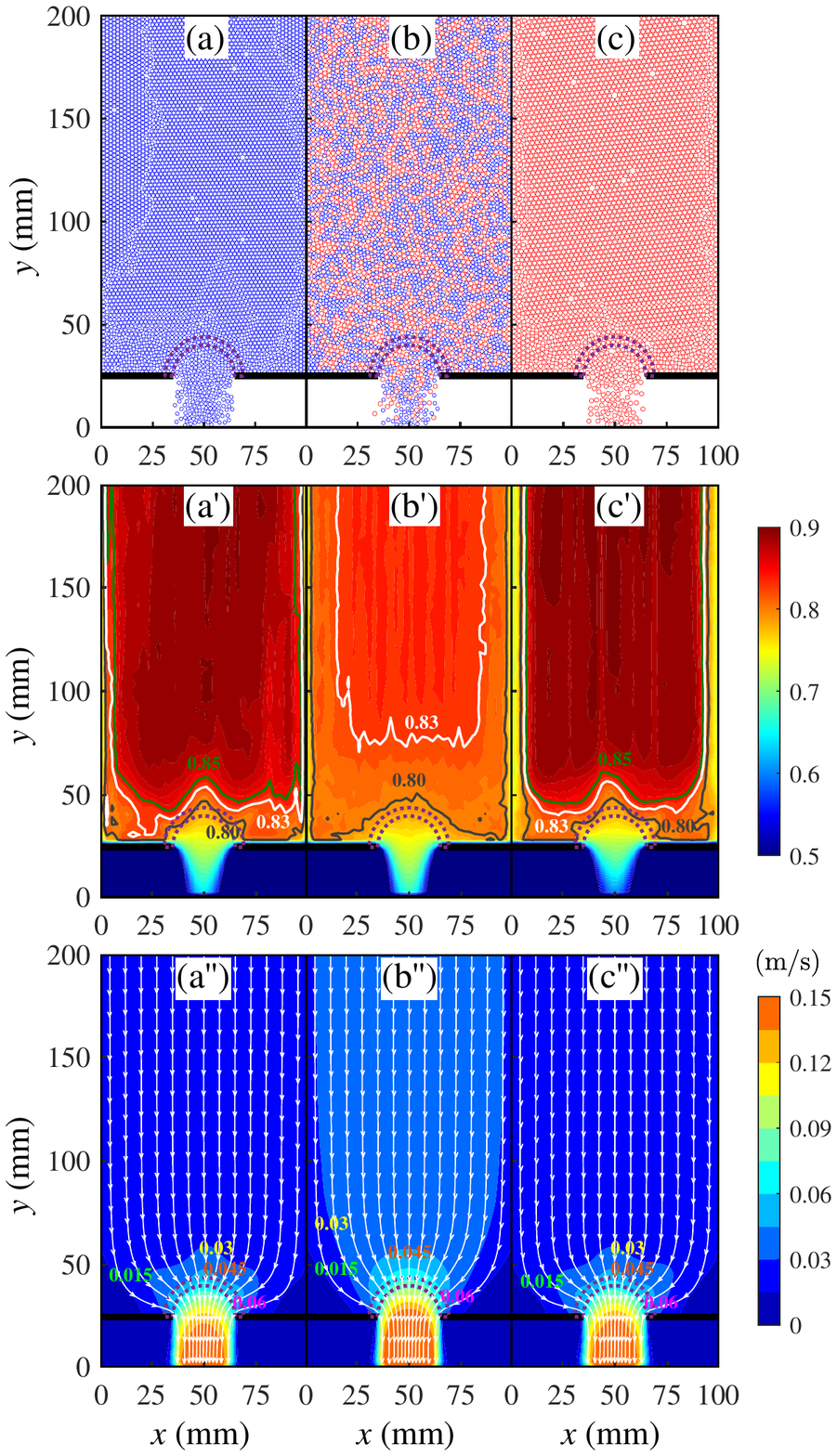}
  \caption{(Color online). (a)(b)(c) Snapshots of steady dense flow. (a')(b')(c') Distribution of dynamical packing fraction. (a'')(b'')(c'') Distribution of grain velocity. The first, second and third columns correspond to the mixing indexes $X_{\rm B}=0.0$, $0.5$ and $1.0$. The diameter ratio is fixed as $\gamma=1.2$. The outlet size is $D=30.0~{\rm mm}$. The key areas are marked with purple dash lines in all figures.}
\label{fig:FigPackDist}
\end{figure}

Fig.\ref{fig:FigQoutXB} plots the time-averaged flow rate as a function of the mixing index for two outlet sizes $D=20~{\rm mm}$ and $30~{\rm mm}$. A non-monotonic behavior is observed for diameter ratios $\gamma=1.2,1.5$ and $2.0$. Compared with the monodisperse flow with small grains, adding a small amount of big grains will obviously increase the flow rate. However, the flow rate decreases after the addition of big grains cross a specific threshold. In general, granular flowability is improved when a low amount of small grains is added into the monodisperse flow with big grains, and the maximum flow rate is observed at a certain low mixing index.

Given that flow rate is determined by the physical quantities of grain velocity and packing fraction, the detailed variances in flow rate resulting from the mixing degree of bidisperse grains warrant further investigation. The initial static packing fraction is examined before the start of granular flow. Even though the grains are randomly poured into the hopper, the size monotonicity results in near crystallization everywhere for the monodisperse grains. A higher static packing fraction $\phi=0.87$ is obtained for $X_{\rm B}=0.0$ and $1.0$. When the grains are bidisperse, the ordered structure is destroyed and the grains randomly pile up in the hopper. The static packing fraction is $\phi=0.83$ for $X_{\rm B}=0.5$.

Fig.\ref{fig:FigPackDist}(a)(b)(c) present the snapshots of steady dense flows for the same conditions shown in Fig.\ref{fig:FigForceDist}. Compared with initial static packing, due to their monodisperse size, the interior crystallization structure of the grains is kept at the region located far from the outlet for the monodisperse flow ($X_{\rm B}=0.0$ and $1.0$) shown in Fig.\ref{fig:FigPackDist}(a)(c). The close-packed structure becomes loose when the grains approach the outlet. By contrast, an ordered arrangement is never observed for the bidisperse flow ($X_{\rm B}=0.5$) from the beginning of the granular flow to the stable flow state as shown in Fig.\ref{fig:FigPackDist}(b). Fig.\ref{fig:FigPackDist}(a')(b')(c') plots the corresponding distribution of dynamical packing fraction in the steady dense flow state. The dynamical packing fraction at the region located far from the outlet is always higher than that around the outlet. A half circular-like shape appears around the outlet and almost overlaps with the key area. The dynamical packing fraction in the key area as denoted by the dotted lines is $\phi=0.80$ for $X_{\rm B}=0.0$, $1.0$ and $\phi=0.78$ for $X_{\rm B}=0.5$. After the grains pass through the key area, the packing fraction decreases continuously until the grains flow out of the hopper for both monodisperse and bidisperse flows.

Fig.\ref{fig:FigPackDist}(a'')(b'')(c'') plots the grain velocity distribution using the same parameters in Fig.\ref{fig:FigForceDist}. The streamlines of granular flow are represented by the white solid lines. The contour color denotes the magnitude of grain velocity. Two stagnant zones are located at the left and right walls as denoted by $0.015~{\rm m/s}$ for each case. The streamlines are a series of parallel lines that emerge when the grains are located far from the outlet and then converge as these grains approach the outlet. The streamlines for the bidisperse case $X_{\rm B}=0.5$ are slightly more deeper than those for the monodisperse cases $X_{\rm B}=0.0$ and $1.0$. The region located far from the outlet is occupied by different colors, where blue corresponds to $X_{\rm B}=0.0$ and $1.0$ and cyan corresponds to $X_{\rm B}=0.5$ as labelled by $0.03~{\rm m/s}$ and $0.045~{\rm m/s}$ respectively. The ordered and disordered arrangements of grains produce two flow patterns, namely, mass flow and funnel flow, which will be further explored in the following section. A similar half circular-like shape reappears around the outlet, and overlaps with the key area as indicated by the dotted lines. The corresponding grain velocity is labelled as $0.06~{\rm m/s}$.
\begin{figure}[htbp]
  \centering
  \includegraphics[width=7 cm,trim=115 280 115 250,clip]{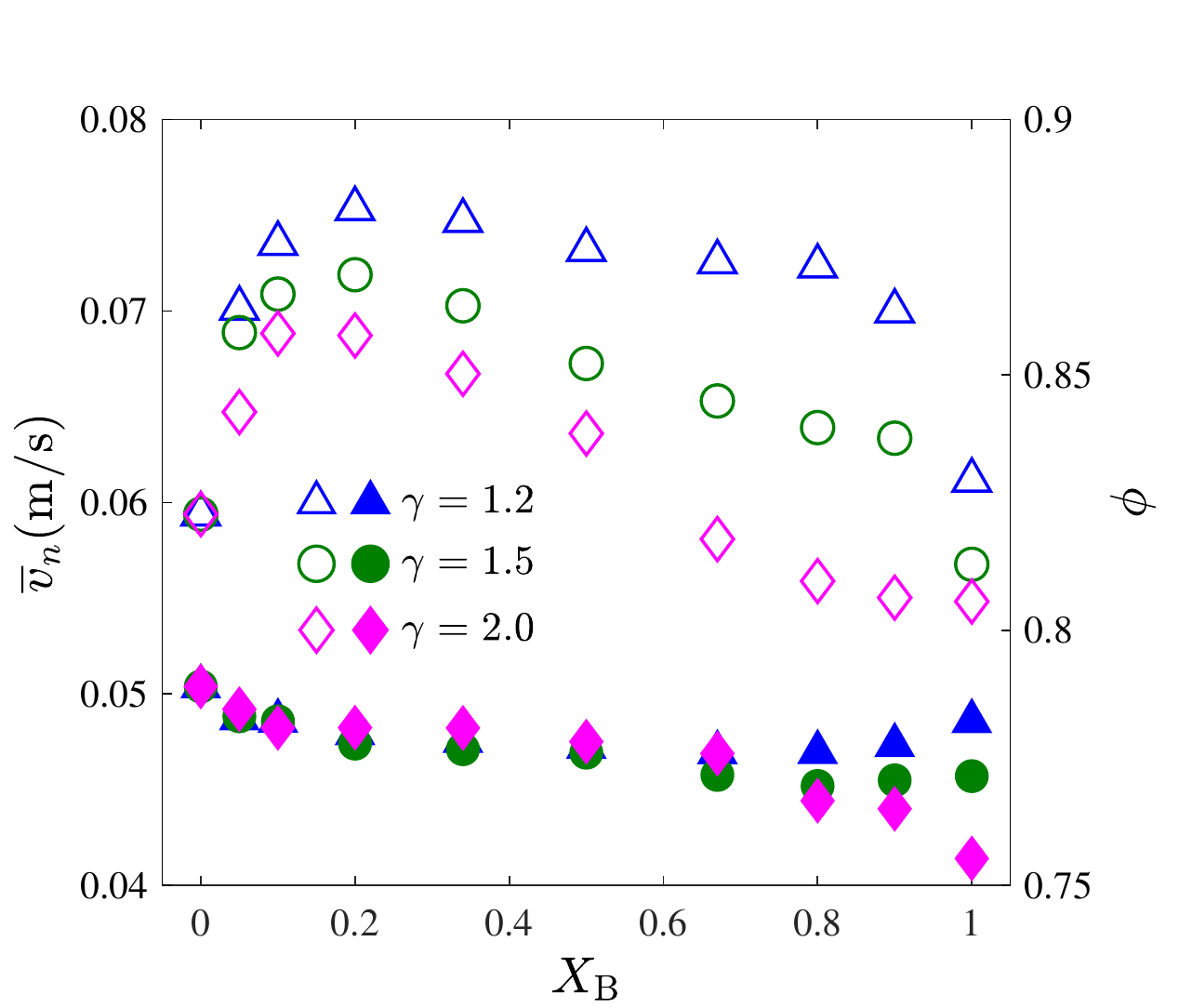}
  \caption{(Color online). The mean normal velocity ${\overline v}_n$ (open symbols) and local packing fraction (solid symbols) in the key area versus the mixing index $X_{\rm B}$. The diameter ratios $\gamma=1.2$, $1.5$, and $2.0$ are used. The outlet size is $D=30.0 {\rm mm}$.}
\label{fig:FigVnPackKey}
\end{figure}

Simulation results indicate that the flow properties are closely related to the local flow characteristics around the outlet, i.e., key area. We now examine how the mixing degree affects the two decisive physical quantities of flow rate in the key area. In Fig.\ref{fig:FigVnPackKey}, the mean normal grain velocity and packing fraction of the key area are plotted as functions of the mixing index for $\gamma=1.2,1.5$ and $2.0$. The outlet size is fixed at $D=30~{\rm mm}$. The maximum normal grain velocity is achieved for the bidisperse flow with smaller $X_{\rm B}$. Meanwhile, the monodisperse flow with small grains always has the  largest packing fraction. The variance in grain velocity is consistent with that in flow rate. Simulations are also performed for outlet size $D=20~{\rm mm}$. The same results are obtained, that is, adding a low amount of big grains into the monodisperse flow with small grains will increase of the grain velocity and reduce the packing fraction, thereby suggesting that increasing grain velocity can effectively improve flow rate \cite{Huang2011PLA}.
\begin{figure*}[htbp]
\centering
\includegraphics[width=0.9\textwidth,trim=20 340 25 330,clip,angle=0]{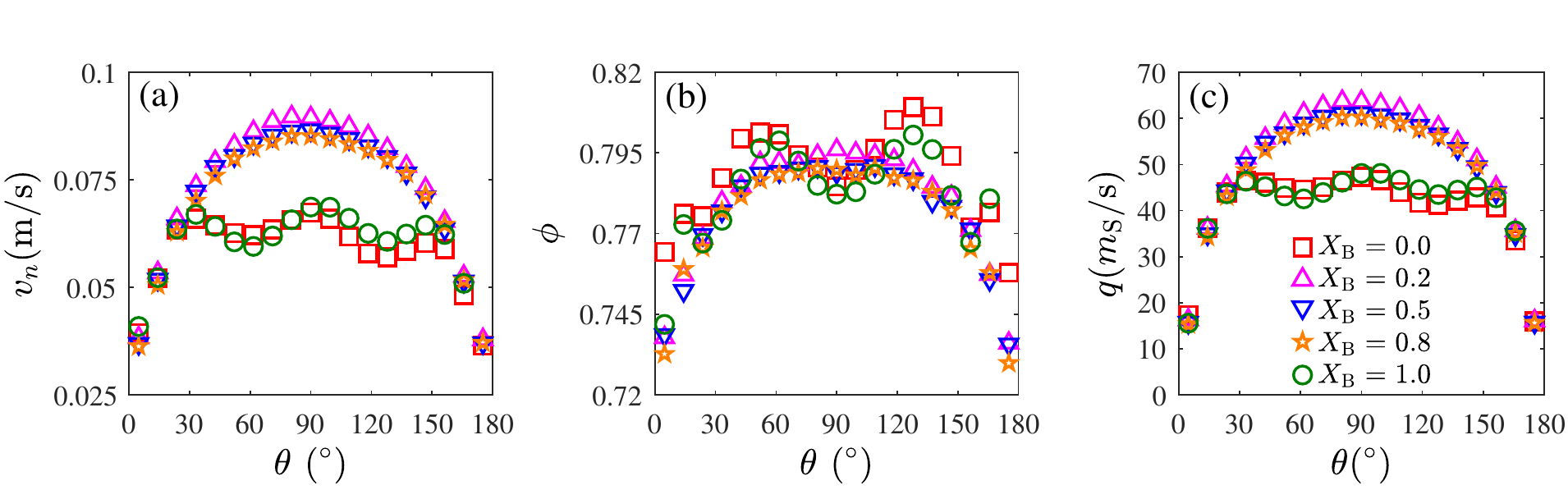}
  \caption{(Color online). Normal velocity $v_n$, packing fraction $\phi$, and flow rate $q$ profiles of the subregions in the key area as a function of the angle $\theta$ shown in Fig.\ref{fig:FigModelKey}(c). The diameter ratio is fixed as $\gamma=1.2$.}
\label{fig:FigVnPackQTheta}
\end{figure*}

Fig.\ref{fig:FigVnPackQTheta} shows the time-averaged normal velocity $v_{n}$, packing fraction $\phi$ and flow rate $q$ in the subregions of the key area for a fixed diameter ratio $\gamma=1.2$. Both the monodisperse flows ($X_{\rm B}=0,1.0$) and the bidisperse flows ($X_{\rm B}=0.2,0.5,0.8$) share the same symmetrical profile. In Fig.\ref{fig:FigVnPackQTheta}(a), all grain velocities sumultaneously fall together at the two side regions. The grain velocity in the center subregions for the monodisperse flow is always smaller than that for the bidisperse flow. Furthermore, the former has a fluctuating shape, whereas the latter has a parabolic shape. In Fig.\ref{fig:FigVnPackQTheta}(b), the monodisperse flow always has a higher packing fractions than that of the bidisperse flow except at the center regions. However, the fluctuating and parabolic distributions are preserved for monodisperse and bidisperse flows, respectively, thereby suggesting that the ordered crystallization structure leads to fluctuations in grain velocity and packing fraction in the monodisperse flow. A symmetrical distribution is also observed for the flow rate shown in Fig.\ref{fig:FigVnPackQTheta}(c). Fig.\ref{fig:FigVnPackQTheta} proves that the normal velocity mainly determines the flow rate in the subregion and that the fluctuation in packing density play a dominant role in flow rate  fluctuation \cite{Huang2018CPB}.
\begin{figure*}[htbp]
\centering
\includegraphics[width=12 cm,trim=60 210 60 220,clip]{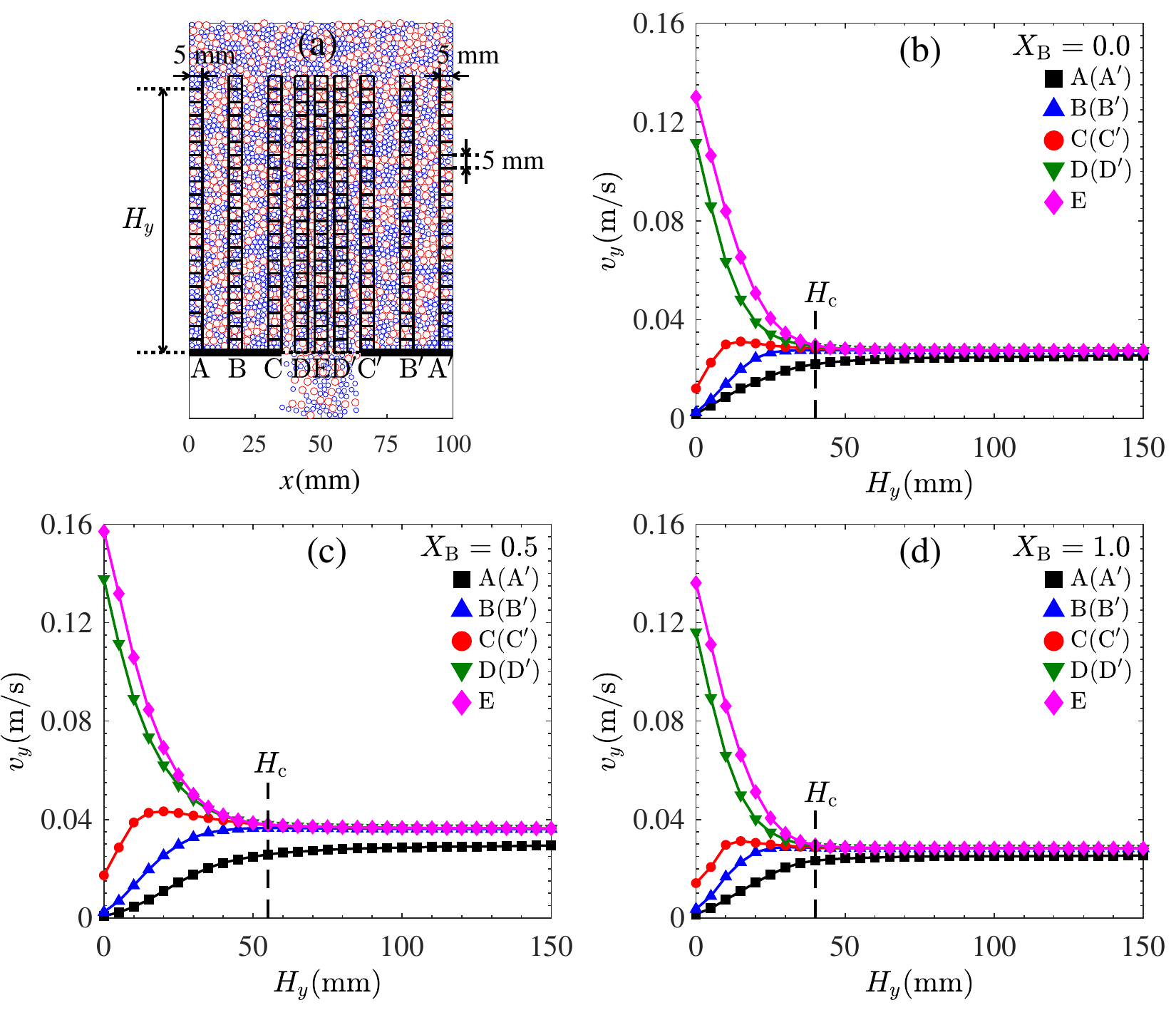}
  \caption{(Color online). (a) Snapshot of the granular vertical velocity extraction area; (b)(c)(d) Mean vertical velocity of selected areas as a function of height for mixing indexes $X_{B}=0.0,0.5$ and $1.0$, respectively. The outlet size is fixed as $D=30.0~{\rm mm}$. The diameter ratio is set as $\gamma=1.2$.}
\label{fig:FigVyHd}
\end{figure*}

\begin{figure}[htbp]
\centering
  \includegraphics[width=7 cm,trim=140 290 140 290,clip]{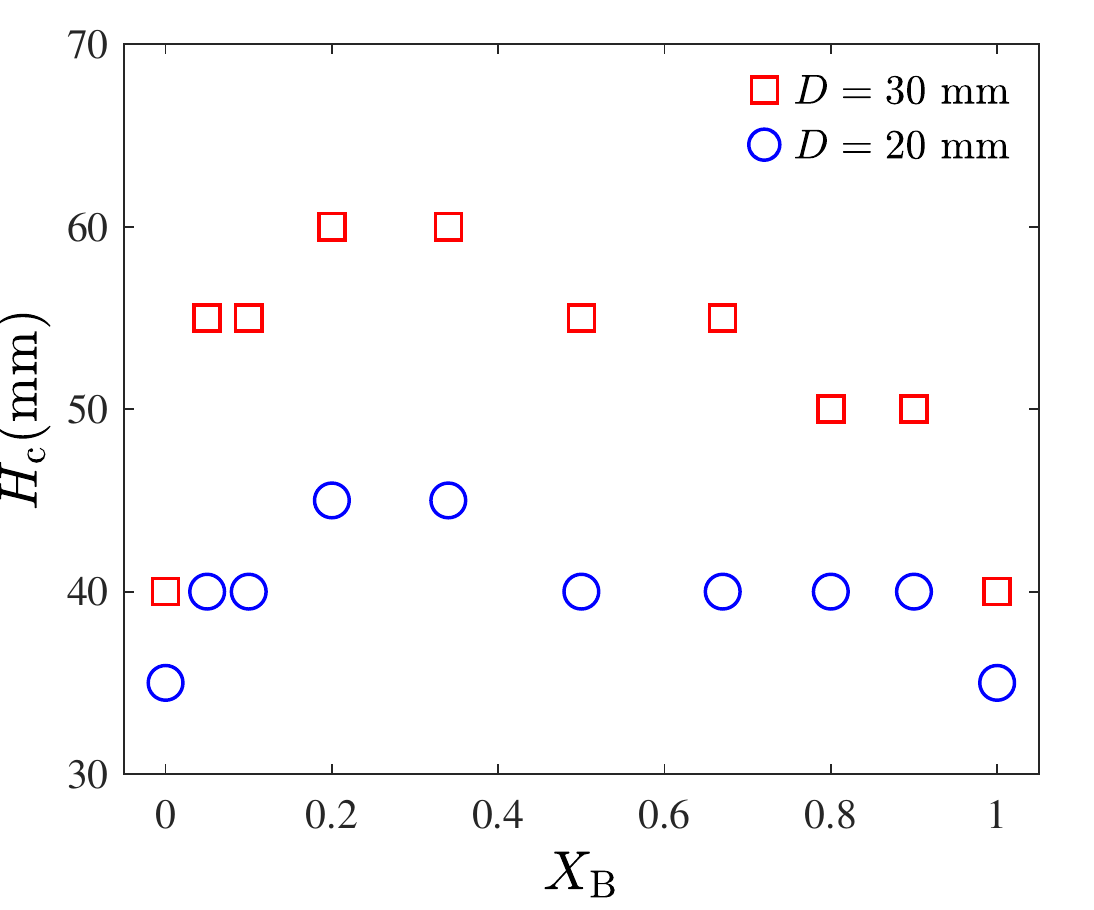}
  \caption{(Color online). Relationship between the critical transition height and the mixing index. The diameter ratio is fixed as $\gamma=1.2$.}
\label{fig:FigHtXB}
\end{figure}

According to the above observations, the monodisperse or bidisperse grains affect on the flow behaviors because grain dispersity leads to an ordered and disordered interior packing structure. To illustrate the detailed influence of mixing degree on the interior flow patterns, i.e., mass flow and funnel flow, the vertical mean velocities ($v_{y}$) of the grains located at positions $A(A'), B(B'), C(C'), D(D')$ and $E$ are measured as shown in Fig.\ref{fig:FigVyHd}(a). The selected regions have a width of $5~{\rm mm}$ and height of $150~{\rm mm}$. Each square has a height of $5~{\rm mm}$. The intersection between the grain and cell is calculated at every time step. The vertical mean velocity as a function of the granular packing height ($H_{y}$) is presented in Fig.\ref{fig:FigVyHd}(b), (c) and (d) for $X_{\rm B}=0,0.5$ and $1.0$, respectively. Given the symmetry, the results at the positions of $A(A'), B(B'), C(C')$ and $D(D')$ are averaged. The critical transition height $H_{\rm c}$ is introduced to decide the flow pattern transition between the mass and funnel flows where the vertical grain velocity coincides at the positions of $B$ and $E$. When $H_{y}>H_{\rm c}$, except those near the side walls, the grains demonstrate a uniform vertical velocity, which suggests that they flow down in the mass flow pattern. When $H_{y}<H_{\rm c}$, the flow takes a funnel flow pattern until the grains flow across the outlet. The grain velocity continues to increase at positions $D$ and $E$ when approaching the outlet, and slows down to zero at the positions $A$, $B$, and $C$ given the presence of baffles.

Fig.\ref{fig:FigHtXB} plots the critical transition height $H_{\rm c}$ as a function of mixing index for outlet sizes $D=20$ and $30~{\rm mm}$. The variance trend of $H_{\rm c}$ is similar to those of flow rate $Q$ and mean normal velocity ${\overline v}_n$ as shown in Fig.\ref{fig:FigQoutXB} and Fig.\ref{fig:FigVnPackKey}. The critical transition height for the monodisperse flow is obviously smaller than that for the bidisperse flow. The maximum height is reported at a low mixing index. The interior arrangement of the grains has an inherent and robust influence on the flow properties. Moreover, adding a low amount of big grains into the monodisperse flow having small grains can effectively increase the grain velocity, especially around the outlet, compared with the monodisperse flows having only small or big grains. This results may be ascribe to the breakdown of the ordered packing structure, which increase the tendency for a mass-to-funnel flow transition to occur in the hopper.

\section{Conclusions}

This work numerically examines the bidisperse dense granular flow in a two-dimensional hopper by using the discrete element method. The analysis of the contact force between grains reveals the existence of a half-circular dynamical force arch above the outlet. Although the size of the force arch contradicts the free-fall arch hypothesis, the simulation results validate Beverloo's law that describes the relationship between flow rate and outlet size. Flow rate has a non-monotonic dependence on the mixing index and reaches its maximum value at a small mixing index.

The monodisperse flow reveals a crystallized structure before the start of the granular flow. While, grains pile up in a disorderly manner in the bidisperse case which corresponds to a smaller packing fraction. When the flow is triggered, the initial close packing becomes loose especially around the outlet. The analysis of both packing fraction and grain velocity reveals a significant change in flow behavior at the key area. The ordered and disordered packing structures lead to the fluctuation and parabolic profiles of the normal grain velocity and packing fraction, respectively. Furthermore, the fluctuation in packing density plays a dominant role in the local fluctuation of flow rate, and the parabolic shape of the grain velocity mainly determines the entire profile of flow rate.

The ordered and disordered packing structures for monodisperse and bidisperse flows lead to different critical transition heights $H_{\rm c}$, where the flow transforms from a mass flow pattern to a funnel flow pattern. These results indicate that the earlier occurrence of the  mass-to-funnel flow pattern transition corresponds to higher grain velocity and larger flow rate. These findings provide further insights into the influence of grain bidispersity on the dense flow behaviors. Further studies may test the rheology of bidisperse granular flow by using the continuous model \cite{Macaulay2020,Pouliquen2006}.


\begin{acknowledgments}
This work is financially supported by the National Natural Science Foundation of China (Grant No. 11574153).
\end{acknowledgments}


\begin{thebibliography}{99}
\bibitem[1]{Jaeger1996} H. M. Jaeger, S. R. Nagel, and R. P. Behringer, Granular solids, liquids, and gases, Rev. Mod. Phys. {\bf 68}, 1259 (1996).

\bibitem[2]{Aranson2006} I. S. Aranson, and L. S. Tsimring, Patterns and collective behavior in granular media: Theoretical concepts, Rev. Mod. Phys. {\bf 78}, 641 (2006).

\bibitem[3]{Iverson1997} R. M. Iverson, The physics of debris flows, Rev. Geophys. {\bf 35}, 245 (1997).

\bibitem[4]{PT22P243} R. M. Nedderman and U. T\"{u}z\"{u}n, A kinematic model for the flow of granular materials, Powder Tech. {\bf 22}, 243 (1979).

\bibitem[5]{RobertsCES2002} A. W. Roberts and C. M. Wensrich, Flow dynamics or 'quaking' in gravity discharge from silos, Chem. Eng. Sci. {\bf 57}, 295 (2002).
\bibitem[6]{WojcikPT2012} M. W$\acute{\rm o}$jcik , J. Tejchman and G. G. Enstad, Confined granular flow in silos with inserts - Full-scale experiments, Powder Tech. {\bf 222}, 15 (2012).

\bibitem[7]{ZhangPT2018} Y. Zhang, F. Jia, Y. Zeng, Y. Han, and Y. Xiao, DEM study in the critical height of flow mechanism transition in a conical silo, Powder Tech. {\bf 331}, 98 (2018).

\bibitem[8]{JiShunyingPT2019} S. Ji, S. Wang, and Z. Peng, Influence of external pressure on granular flow in a cylindrical silo based on discrete element method, Powder Tech. {\bf 356}, 702 (2019).

\bibitem[9]{JiShunyingSM2020} S. Wang, M. Zhuravkov, and S. Ji, Granular flow of cylinder-like particles in a cylindrical hopper under external pressure based on DEM simulations, Soft Matter {\bf 16}, 7760 (2020).

\bibitem[10]{HuangEPJE2013} D. C. Huang, M. Lu, S. Sen, M. Sun, Y. D. Feng and A. N. Yang, Spin Brazil-nut effect and its reverse in a rotating double-walled drum, Eur. Phys. J. E {\bf 36}, 9855 (2013).
\bibitem[11]{ShiPRE2007} Q. Shi, G. Sun, M. Y. Hou and K. Q. Lu, Density-driven segregation in vertically vibrated binary granular mixtures, Phys. Rev. E {\bf 75}, 061302 (2007).

\bibitem[12]{WangCPB2017} W. G. Wang, Z. Z. Zhou, J. Zong and M. Y. Hou, DEM simulation of granular segregation in two-compartment system under zero gravity, Chin. Phys. B {\bf 26}, 044501 (2017).

\bibitem[13]{Bursik2005} M. Bursik, A. Patra, E. B. Pitman, C. Nichita, J. L. Macias, R. Saucedo and O. Girina, Advances in studies of dense volcanic granular flows, Rep. Prog. Phys. {\bf 68}, 271 (2005).

\bibitem[14]{Denisov2016} D. V. Denisov, K. A. L\"{o}rinca, J. T. Uhi, K. A. Dahmen and P. Schall, Universality of slip avalanches in flowing granular matter, Nat. Commum. {\bf 7}, 10641 (2016).

\bibitem[15]{Hagen1852} B. P. Tighe and M. Sperl, Pressure and motion of dry sand: translation of Hagen's paper from 1852, Granular Matter {\bf 9}, 141 (2007).
\bibitem[16]{Beverloo1961} W. A. Beverloo, H. A. Leniger, J. van de Velde, The flow of granular solids through orifices, Chem. Eng. Sci. {\bf 15}, 260 (1961).

\bibitem[17]{NeddermanCES1982} R. M. Nedderman, U. T\"{u}z\"{u}n, S. B. Savage, and G. T. Houlsby, The flow of granular materials-I: Discharge rates from hoppers, Chem. Eng. Sci. {\bf 37}, 1597 (1982).

\bibitem[18]{Zuriguel2012PRL108P248001} A. Janda, I. Zuriguel, and D. Maza, Flow rate of particles through apertures obtained from self-Similar density and velocity profiles, Phys. Rev. Lett. {\bf 108}, 248001 (2012).

\bibitem[19]{AguirrePRL2010} M. A. Aguirre, J. G. Grande, A. Calvo, L. A. Pugnaloni, and J.-C. G$\acute{\rm e}$minard, Pressure independence of granular flow through an aperture, Phys. Rev. Lett. {\bf 104}, 238002 (2010).

\bibitem[20]{ZhengPT345P676Y2019} H. W. Zhu, L. P. Wang, Q. F. Shi, L. S. Li, and N. Zheng, Improvement in flow rate through an aperture on a conveyor belt: Effects of bottom wall and packing configurations, Powder Tech. {\bf 345}, 676 (2019).
\bibitem[21]{HuangPT2021} X. Y. Zhou, S. K. Liu, Z. H. Zhao, X. Li, C. H. Li, M. Sun, and D. C. Huang, Dilute-to-dense flow transition and flow-rate behavior of lateral bifurcated granular flow, Powder Tech. {\bf 383}, 536 (2021).

\bibitem[22]{ArteagaCES1990} P. Arteaga and U. T\"{u}z\"{u}n, Flow of binary mixtures of equal-density granules in hoppers - size segregation, flowing density and discharge rates, Chem. Eng. Sci. {\bf 45}, 205 (1990).

\bibitem[23]{HumbyCES1998} S. Humby, U. T\"{u}z\"{u}n, and A. B. Yu, Prediction of hopper discharge rates of binary granular mixtures, Chem. Eng. Sci. {\bf 53}, 483 (1998).

\bibitem[24]{BenyaminePRE2014} M. Benyamine, M. Djermane, B. Dalloz-Dubrujeaud, and P. Aussillous, Discharge flow of a bidisperse granular media from a silo, Phys. Rev. E {\bf 90}, 032201 (2014).

\bibitem[25]{ZhouPRE2015} Y. Zhou, P. Ruyer, and P. Aussillous, Discharge flow of a bidisperse granular media from a silo: Discrete particle simulations, Phys. Rev. E {\bf 92}, 062204 (2015).
\bibitem[26]{Huang2006PRE} D. C. Huang, G. Sun, and K. Q. Lu, Relationship between the flow rate and the packing fraction in the choke area of the two-dimensional granular flow, Phys. Rev. E {\bf 74}, 061306 (2006).

\bibitem[27]{Huang2011PLA} D. C. Huang, G. Sun, and K. Q. Lu, Influence of granule velocity on gravity-driven granular flow, Phys. Lett. A {\bf 375}, 3375 (2011).

\bibitem[28]{Huang2018CPB} T. W. Wang, X. Li, Q. Q. Wu, T. F. Jiao, X. Y. Liu, M. Sun, F. L Hu, and D. C. Huang, Numerical simulations of dense granular flow in a two-dimensional channel: The role of exit position, Chin. Phys. B {\bf 27}, 124704 (2018).

\bibitem[29]{Cundall1979} P. A. Cundall, and O. D. L. Strack, A discrete numerical model for granular assemblies, G$\acute{\rm e}$otechnipue {\bf 29}, 47 (1979).

\bibitem[30]{Schafer1996} J. Sch\"{a}fer, S. Dippel, and D. E Wolf, Force Schemes in Simulations of Granular Materials, J. Phys. I France {\bf 6}, 5 (1996).
\bibitem[31]{Huang2007CTP} D. C. Huang, G. Sun, and K. Q. Lu, A stochastic description of transition between granular flow states, Commun. Theor. Phys. {\bf 48}, 729 (2007).

\bibitem[32]{Macaulay2020} M. Macaulay, and P. Rognon, Two mechanisms of momentum transfer in granular flows, Phys. Rev. E {\bf 101}, 050901(R) (2020).

\bibitem[33]{Pouliquen2006} O. Pouliquen, C. Cassar, P. Jop, Y. Forterre, and M. Nicolas, Flow of dense granular material: towards simple constitutive laws, J. Stat. Mech. {\bf 7}, P07020 (2006).

\end{thebibliography}
\end{document}